# Exploring the effect of varying regime of ion fluence on optical and surface electronic properties of CVD grown graphene

**Tanmay Mahanta[1], Sanjeev Kumar[1], D. Kanjilal[2] and Tanuja Mohanty[1, a)]**

[1]*School of Physical Sciences, Jawaharlal Nehru University, New Delhi - 110067, INDIA.*
[2]*Inter University Accelerator Centre, Aruna Asaf Ali Marg, New Delhi - 110067, INDIA.*

[a)]*Corresponding author: tanujajnu@gmail.com*

## Abstract

In this work, the effect of the ion fluence-dependent defect formation on the modification of surface electronic and optical properties of graphene has been investigated. The chemical vapor deposited (CVD) graphene was irradiated with swift heavy ion (SHI) (70 MeV $Si^{+5}$) beam at different fluence to study the defect formation mechanism and the role of it in modulating its surface electronic property such as work function. At low ion dose, acceptor doping via vacancy creation was observed; whereas dense electronic excitation dominated extended defects was realized at higher dose, which subsequently transforms the crystalline graphene into amorphous carbon. The results from UV-Vis spectroscopy, Raman spectroscopy and scanning Kelvin probe microscopy (SKPM) support the fact. Thus a new pathway of transformation of graphene under SHI irradiation was explored where ion dose could be the main factor to realize several effects in graphene.

# 1. Introduction

Graphene, a single layer of graphite, is predominantly occupying position in the field of science and technology since its discovery due to its unique and outstanding properties [1]. Several novel properties that are not available in other materials at room temperature and ambient conditions have been observed in graphene. The charge carrier in graphene, unlike other known materials, is massless Dirac Fermions [2], which makes the conductivity in graphene the highest at room temperature. In the Brillouin zone around the K point or Dirac point, the energy-momentum relation is linear, responsible for many remarkable properties in graphene [2]. The room temperature quantum Hall effect [3] and the breaking of the Born-Oppenheimer approximation are observed in graphene [4]. It is the thinnest and stiffest material known to date, with a tensile strength of 125 GPa, an elastic modulus of 1.1 TPa, and a two-dimensional ultimate plane strength of 42 N-m$^{-2}$ [5]. It possesses high carrier mobility of $2\times10^5$ cm$^2$ (V$^{-1}$ s$^{-1}$) which gets affected by impurities and defects [6]. Graphene is made out of carbon atom arranged in a hexagonal honeycomb lattice with two atoms in a unit cell. Each layer forms sp$^2$ hybridization to bond with other carbon atoms, and the non-bonding electrons form π bonds that move freely between layers [7].

Despite these novel properties, graphene's use is limited in the nano-electronic device industry as the required finite on-off ratio is not present in pristine form [2]. As the inversion symmetry is preserved, it cannot be used to make piezoelectric devices either [8]. Several efforts have been made to modify graphene's properties and harness it for device applications. Controlled doping and the creation of vacancies are some of these methods followed for this purpose [9, 10]. Ion beam irradiation is a suitable technique by which, in a controlled manner, any material can be doped with donor or acceptor atoms of choice or vacancy can be created in it. By externally controlling incident angle, fluence of the desired ion beam, induced effects on target materials could be controlled. Depending upon the energy of the ion beam, processes like adsorption, ion implantation, track formation, etc., could be realized in the target materials. Therefore, it is essential to investigate the effect of ion beam on graphene, which modifies its physical properties.

In this work, single/bilayers of graphene deposited on $SiO_2/Si$ were irradiated by 70 MeV $Si^{+5}$ ion beam at varying fluence. It is observed that at a fluence of $5\times10^{11}$ ions-cm$^{-2}$, the vibrational and surface electronic properties are modified by the introduction of dopants. The Raman spectra, absorption spectra, and work function mapping are supporting the fact. At high fluence ($1\times10^{12}$ ions-cm$^{-2}$ & $5\times10^{12}$ ions-cm$^{-2}$), a defect-induced tensile strain is imposed in graphene, beyond which complete transformation to amorphous carbon is observed.

## 2. Experimental Details

CVD-grown graphene sheets (single/bilayers) on $SiO_2/Si$ substrate (1cm×1cm dimension) were purchased from Graphene Supermarket. The thickness of $SiO_2$ on Si is 285 nm. The thin film of graphene was initially cleaned with acetone to remove any external elements before being used for further experiments. The 16 MV Pelletron accelerator was employed to irradiate the samples by 70 MeV $Si^{+5}$ beam with fluence $5\times10^{11}$ ions-cm$^{-2}$, $1\times10^{12}$ ions-cm$^{-2}$, $5\times10^{12}$ ions-cm$^{-2}$ & $5\times10^{13}$ ions-cm$^{-2}$ at normal incidence. The pressure inside the chamber was $1.33\times10^{-9}$ bar. PANlytical X'pert PRO (Cu-K$_\alpha$ line) was used for XRD studies of graphene. For optical studies of pristine and irradiated samples, Raman analysis was done by Witec Alpha 300 (using 532 nm Laser, 100 μW) and UV-VIS spectroscopy was done using Shimadzu UV 2600 UV-Vis Spectrophotometer in reflectance mode. The corresponding absorption data was extracted using the Kubelka Monk transformation method available in the setup. The work function of graphene was measured in terms of contact potential difference (CPD) using SKPM of KP Technology.

## 3. Results
## 3.1 XRD Studies

The XRD plot of the pristine sample is shown in Fig.1. The characteristic peak of graphene [11, 12] is observed at an angle ~24º. The sharpness of the peak is less, which may be due to the existence of single/bilayers of graphene and discontinuity in the graphene film. No other peak corresponding to different planes other than (002) is observed, implying that graphene's (002) plane is abundant in our sample.

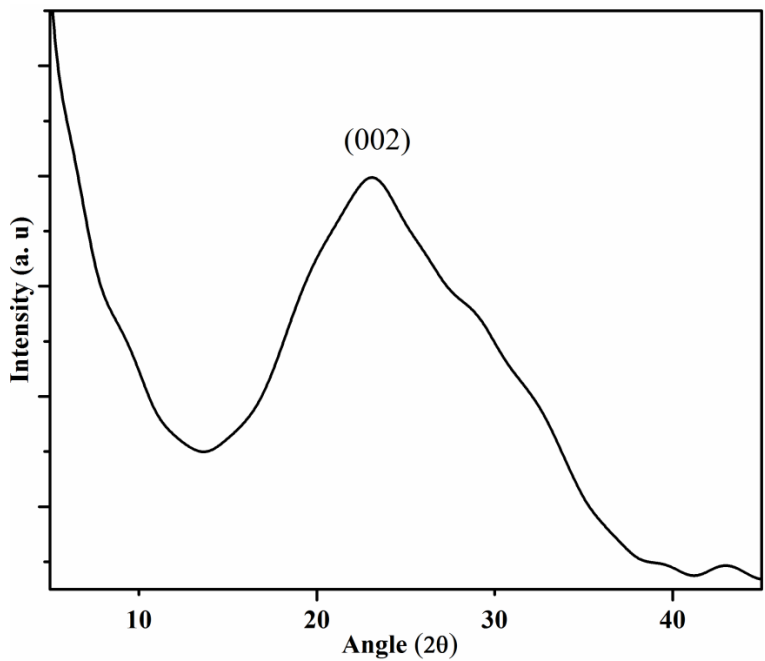

**Fig. 1.** XRD plot of pristine graphene. The peak at ~24º corresponds to the (002) plane.

## 3.2 Raman studies

The graphene is identified by the Raman spectrum indirectly through its signature peaks. A sharp 2D peak caused by the vibration of the hexagonal ring of carbon is observed in graphene; the Kohn anomaly is the main contributor to its strength [13]. For a single layer of graphene, it is as much as four times the strength of the G band, which is associated with the doubly degenerate in-plane transverse optical phonon mode and longitudinal optical phonon mode at $\Gamma$ point (iTO & LO) [13]. The presence of defects like doping, track formation, vacancy generation, etc., affects the intensity of these two Raman peaks. The appearance of new peaks, like, D and D' which quantitatively gives the degree of disorder in the sample, occurs due to the introduction of defects. [13]. The inter-valley double resonance process involving defects and iTO phonons and intra-valley double resonance process involving LO phonons and defects are the contributors of the peaks, respectively.

The Raman spectra of pristine and irradiated samples are shown in Fig. 2.

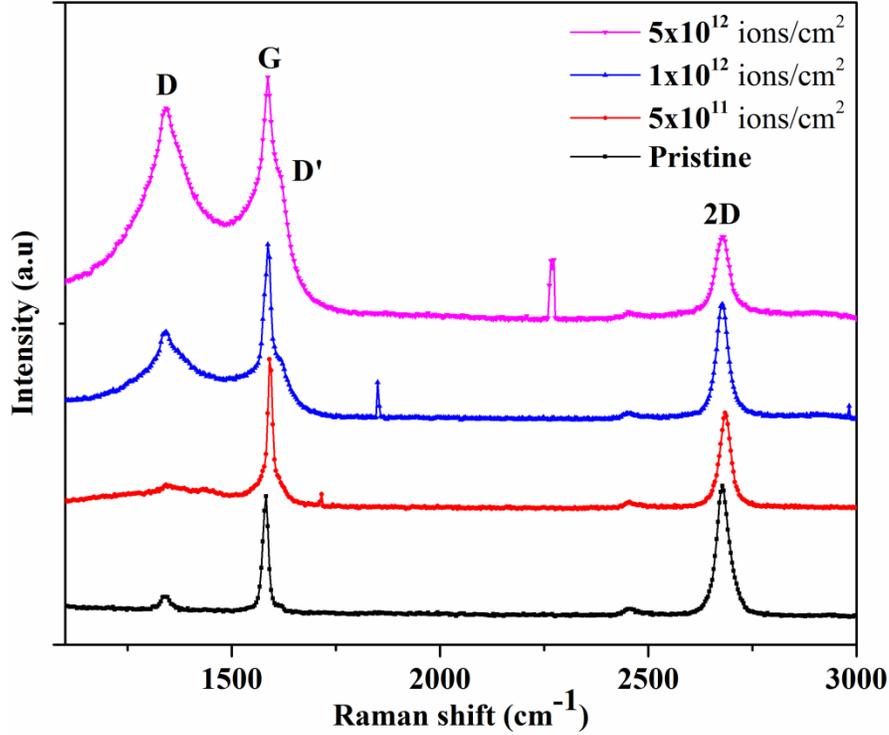

**Fig. 2.** Raman spectra of pristine and Si $^{+5}$ ion irradiated graphene.

The two prominent characteristic peaks of pristine graphene/SiO$_2$/Si viz., 2D and G are observed at positions ~2681.4 cm$^{-1}$ and 1581.25 cm$^{-1}$, respectively (Fig. 2). However, in the pristine sample, the peak position is slightly redshifted from as reported for freestanding, un-doped graphene (G position at 1580 cm$^{-1}$; 2D position at 2690 cm$^{-1}$). The redshift appears due to the effective strain imposed on the graphene layer after depositing on SiO$_2$/Si substrate. Here, the 2D and G peak ratio is slightly higher than 1, implying the presence of single domains with bi-layers of graphene mostly. With increasing ion fluence during ion beam irradiation, the position and intensity of Raman peaks go through changes in the following manner.

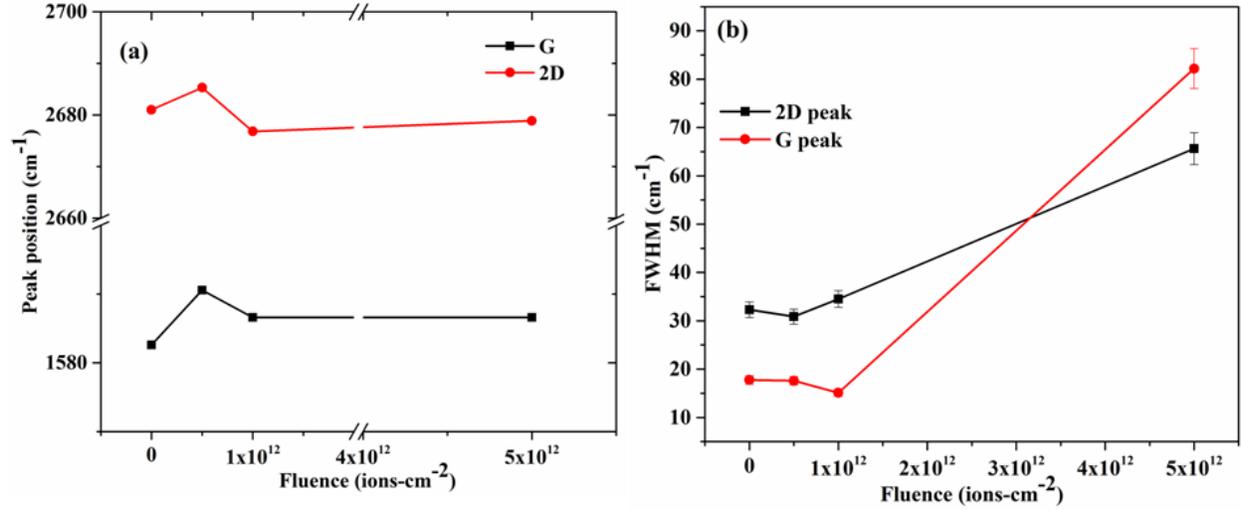

**Fig. 3.** (a) 2D and G Raman peak position vs. fluence in pristine and irradiated graphene. (b) FWHM of 2D and G peaks vs. fluence.

The peak positions (Fig. 3a) of both the peaks are blue-shifted at initial fluence of irradiation. The FWHM (Fig. 3b) is decreased for the ion dose $5\times10^{11}$ ions-cm$^{-2}$. It is reported that doping shifts Raman peak to a lower wavelength, i.e., blueshifts Raman peak (2D peak blueshifts for p doping and redshifts for n doping whereas G peak in both cases gets blueshifted) occurs [14-17]. In the presence of dopants, the effective lifetime of a phonon increases as it shifts the Fermi level; correspondingly, the width of Raman spectrum decreases [14, 18]. For further confirmation, the slope of pos(2D) and pos(G) was taken from different spots of the samples. It was observed that the slope for the sample irradiated with ion dose $5\times10^{11}$ ions-cm$^{-2}$ is ~0.704, which is approximately the ratio assigned for hole doping as documented in the literature [19]. Hence, it can be concluded that the hole doping effect dominates in the sample irradiated with fluence $5\times10^{11}$ ions-cm$^{-2}$.

On irradiation of ion doses higher than $5\times10^{11}$ ions-cm$^{-2}$, FWHM starts increasing, indicating a deviation from pristine graphene structure. The peak positions are redshifted, and the defect-induced $I_D$ peak intensity is increased. As the 2D' peak position of graphene is not affected by doping or charge impurity but strain [20]; we have plotted 2D´ peak with ion fluence to get insight into the effect of ion fluence in imposing strain in graphene. Upon fitting 2D´ peak, a redshift is observed (Fig. 4b), which is a sign of tensile strain produced in graphene.

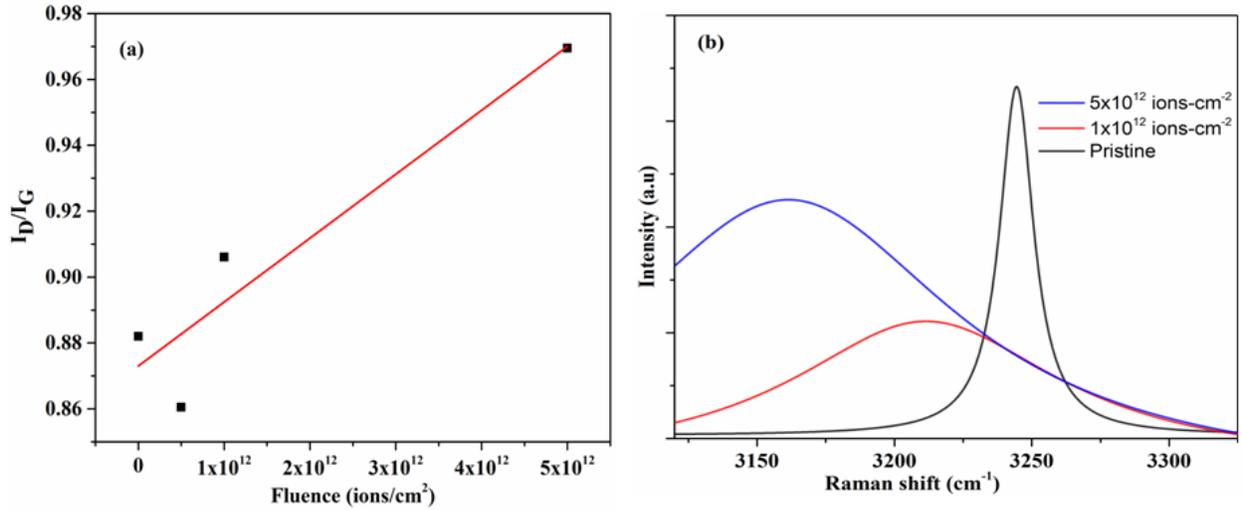

**Fig. 4.** (a) $I_D/I_G$ vs. fluence graph; (b) 2D´ position with increasing fluence.

To calculate the value of induced strain, we have used equation-1, which includes position shifting of 2D and G peaks and original peak positions of the unstrained sample [21].

$$\frac{\Delta\omega}{\omega} = -\gamma_m Tr\epsilon_{ij} \tag{1}$$

Where $\Delta\omega$ is the Raman shift, $\omega$ is Raman peak position in pristine graphene, $\gamma_m$ is Gruneisen parameter, and $\epsilon_{ij}$ is strain tensor. We have focused on 2D peak position to calculate strain as it is more affected by strain than G peak [22]. The analysis reveals that the strain is tensile in nature, with values 0.15% and 0.19% for $1\times10^{12}$ ions/cm² and $5\times10^{12}$ ions/cm² of fluence respectively.

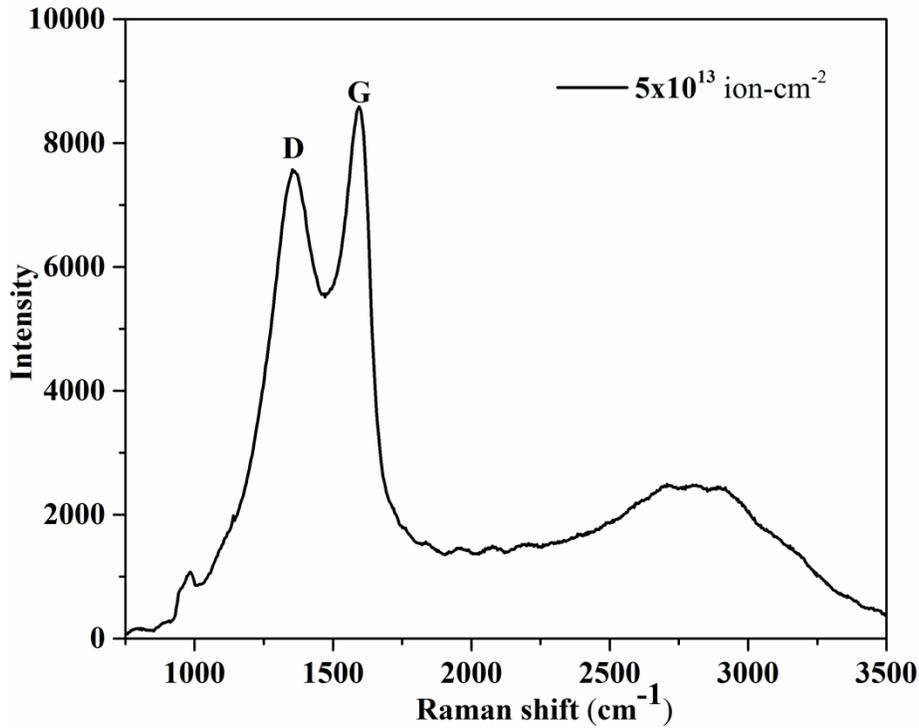

**Fig. 5.** Raman spectrum of $Si^{+5}$ beam irradiated graphene with fluence $5\times10^{13}$ ions-cm$^{-2}$; complete transformation of graphene into amorphous carbon is visible.

The effect of $Si^{+5}$ ion beam at fluence $5\times10^{13}$ ions-cm$^{-2}$ was also investigated. In the corresponding Raman spectrum, the D and G peak is visible with very high intensity; the 2D peak intensity has almost been diminished. These indications imply the complete transformation of graphene film into amorphous carbon (Fig. 5) [23].

## 3.3 Absorbance spectra studies

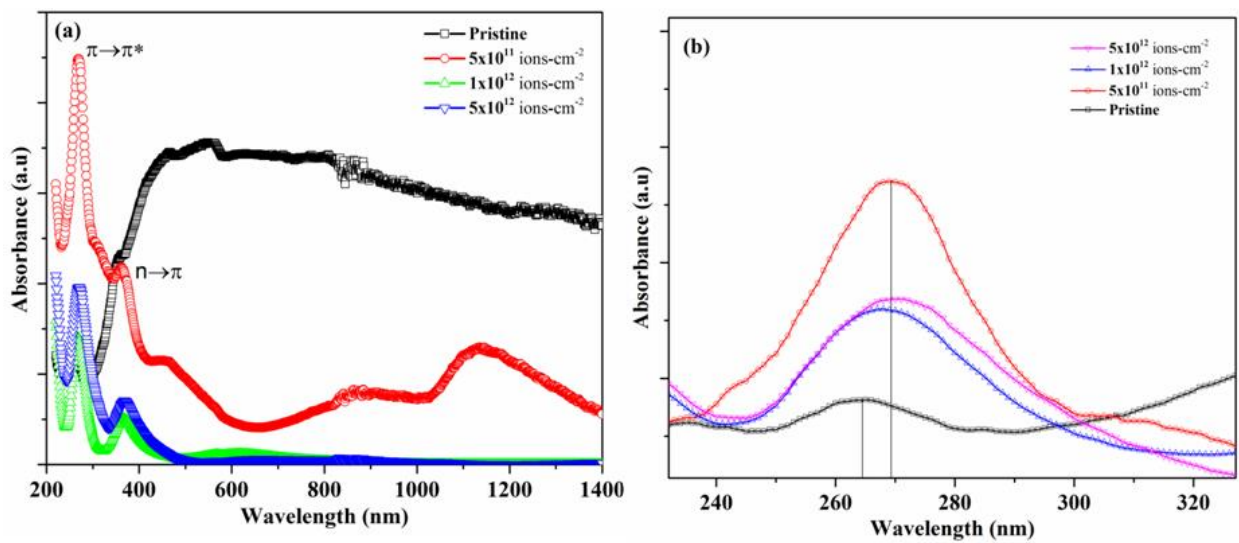

**Fig. 6.** (a) Absorbance spectra of pristine and irradiated graphene. (b) Zoomed view of the peak at ~265 nm.

The absorption spectra of pristine and irradiated graphene is shown in Fig. 6a and in 6b close view of Absorption peak at ~ 265 nm is given. This absorption peak in the UV range could be a signature of the presence of graphene. Graphene consists of carbon atoms in $sp^2$ hybridized state. The presence of the C=C bond in graphene makes the electronic transition from π→π* bonding and anti-bonding state [24]. We have observed a peak at around 265 nm (Fig. 6a) in the pristine sample, connected to π→π* transition of C=C bond. After ion irradiation, at fluence $5\times10^{11}$ ions-$cm^{-2}$, the intensity is highly increased; this may be due to acceptor doping confirmed from Raman spectra, intensifying the absorbance peak [25, 26]. At the same time, the redshift of peak position is observed, which implies the enhancement of screening for which the electron-hole interaction is reduced, and the resonant range of excitonic state is smaller than that of pristine graphene [27, 28]. It is a clear sign of doping in graphene. With the further increment of fluence, the intensity goes higher than the pristine one, which may be due to the presence of damaged sites that absorb light very much [29].

## 3.4 WF Measurement

The work function of graphene on SiO$_2$/Si was calculated using scanning Kelvin probe microscopy (SKPM), which allows us to map the contact potential difference (CPD) between the gold tip embedded in it and the sample. Once the value of CPD is known, the work function value can be estimated using the following equation [30].

$$eV_{CPD} = \phi_{tip} - \phi_{sample} \qquad (2)$$

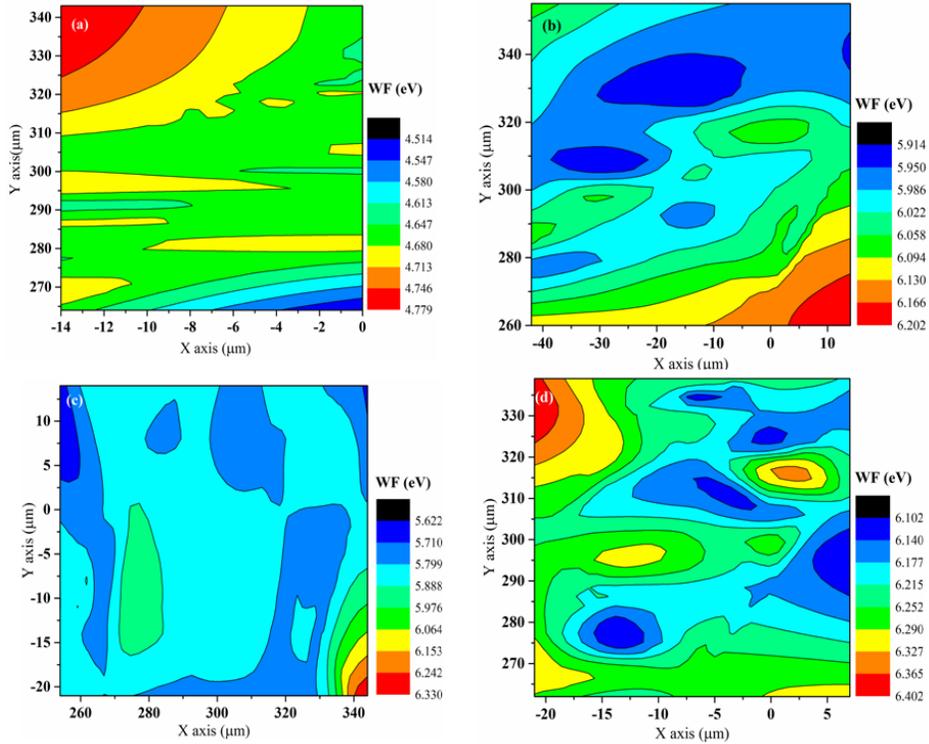

**Fig. 7.** Work function mapping of ion irradiated graphene. (a) pristine, (b) 5×10$^{11}$ ions-cm$^{-2}$, (c) 1×10$^{12}$ ions-cm$^{-2}$, (d) 5×10$^{12}$ ions-cm$^{-2}$.

Where $e$ is the electronic charge, $\phi_{tip}$ and $\phi_{sample}$ are the work function of the gold tip and the sample, respectively. The value of $\phi_{tip}$ is ~5.1 eV. From this information and the value of CPD, the work function of pristine graphene (Fig. 7a) comes out to be ~4.8 eV, which is in good agreement with previously reported results [31]. Now for the sample irradiated with 5×10¹¹ ions-cm⁻², it increases to ~6.06 eV (Fig. 8).

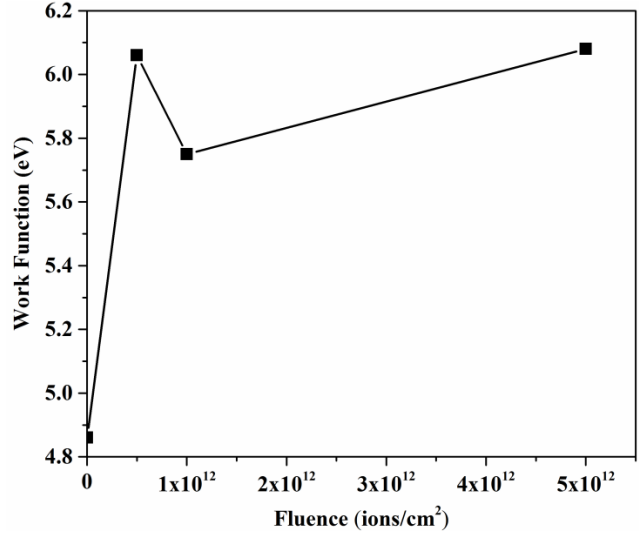

**Fig. 8.** WF evolution of graphene under ion irradiation.

The significant enhancement of the work function of the sample irradiated with fluence 5×10¹¹ ions-cm⁻² is attributed to acceptor doping; where electrons transfer from graphene resulting shifting of Fermi level from Dirac point to valence band. For the samples irradiated beyond this fluence, the defect-induced tensile strain could be the possible reason behind the increment of work function [32]. The doping concentration is calculated using equation-3.

$$n = \frac{1}{\pi}[\Delta E_F / h \, v_F]^2 \qquad (3)$$

Where $v_F$ is the Fermi velocity, it has a value 1×10⁶ *m/s*. $\Delta E_F$ is shifting in WF value, $h$ is Plank's constant. The value of un-doped graphene's work function is ~4.57 eV [33]. It is observed that initially, p doped graphene with concentration 9.63×10¹⁰ cm⁻² increases to 3.86×10¹² cm⁻². The defect-induced tensile strain imposed in the rest samples enhances the WF in graphene as it shifts the Fermi level downwards while keeping the vacuum level almost unaltered [34].

## 4. Discussion

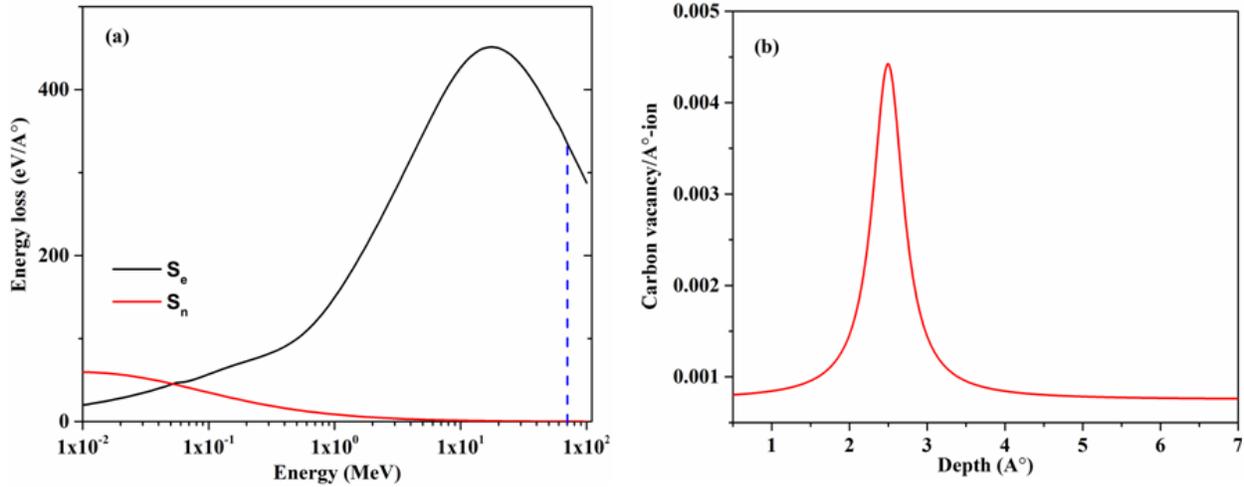

**Fig. 9.** (a) Electronic energy loss and nuclear energy loss of $Si^{+5}$ ions in graphene. (b) Carbon vacancies in graphene.

The ion beam irradiation on graphene results in a significant change in its electronic and vibrational properties. It is expected that the initial fluence of ion beam results in doping of the sample with acceptor atoms, and upon increasing ion fluence, external tensile strain plays major role. It is improbable that a high energetic ion beam with energy as many as 70 MeV will get implanted in the target material directly. It is reported that the surface track created by ion beam irradiation on graphene at glancing angle deposition works as a way of doping graphene from the substrate atoms [35]. It is also reported by Kim et al that the charge transfer between ion beam-induced broken carbon bonds and ambient $O_2$ molecules results in p doping [14]. However, in this work, we have observed that the doping effect neither comes abruptly from the surface tracks nor the ambient oxygen; instead, ion fluence plays an important role here. The possibility of oxygen contamination during ion irradiation is improbable since high vacuum was used throughout the experiment. Nonetheless, had the ambient oxygen doped graphene making bonds at damage sites during characterization, the doping would have been evident in all irradiated samples that was not observed. Since the ion beam irradiation had been taken at normal incidence, the presence of surface tracks at different spots on graphene was not possible. In this case, we have observed a mixed effect of nuclear and electronic energy loss. At low fluence, it creates carbon vacancies in graphene mostly, leading to acceptor doping [36]. The Fermi level downshifts through valence band minima, resulting in available energy levels for electronic transition. The redshift in absorbance spectra and the enhancement of work function can be attributed to that [37]. At higher fluence, tensile strain is imposed due to dense electronic excitation leading to localized damage which increases the defect density thus enhancing the Raman FWHM. We have also investigated the possibility of doping via surface sputtering, but the yield is low (~0.005/ion) and cannot justify the presence of dopants only in a sample irradiated with $5\times10^{11}$ ions-cm$^{-2}$. The electronic energy loss, nuclear energy loss, and carbon vacancy graphs are shown in Fig. 9; where the electronic energy loss is dominant, the nuclear energy loss is low yet at low fluence, can create few vacancies in graphene. The carbon vacancy graph shows that the vacancy is highest at the surface region. Beyond the fluence $5\times10^{11}$ ions-cm$^{-2}$, extended defects in graphene due to dense electronic excitation imposed strain. These graphs justify the mixed effects of ion beam; at low fluence, the dominance of acceptor doping while at higher fluence extended defect induced tensile strain.

## 5. Conclusions

In summary, CVD grown graphene was irradiated with different ion fluence of $Si^{+5}$ ions, and their effect on its optoelectronic properties were investigated. At low fluence ($5\times10^{11}$ ions-cm$^{-2}$), it is observed that graphene is doped with vacancy-induced acceptor dopants, whereas at higher fluence ($1\times10^{12}$ ions-cm$^{-2}$, $5\times10^{12}$ ions-cm$^{-2}$), tensile strain is imposed by localized defect sites. This behavior of CVD graphene can help in tailoring its properties in a controlled way for the intended purpose in optoelectronic devices. The tuned work function of graphene is often desirable in the FET and catalytic applications, where work function and surface charge distribution play a significant role.

## Acknowledgments

The authors are thankful to AIRF, JNU for Raman, XRD measurements; Dr. Supriya Sabbani, SPS, JNU for UV-Vis spectroscopic measurement; IUAC, New Delhi for beam time.